\documentclass[a4paper,11pt]{article}

\usepackage{spstyle}
\usepackage{subfig}
\usepackage{xcolor}
\newenvironment{report}
{\begingroup\color{black}}  
{\endgroup}               

\def\Cwebgen{\texttt{CWebGen}}

\begin{document}
\begin{titlepage}
\begin{center}
	{{\huge CWebGen}}\\ \bigskip
		 {\Large A tool to study colour structure of scattering amplitudes in IR limit} \bigskip \\
	{\large Neelima Agarwal\,, Sourav Pal\,$^a$, Aditya Srivastav\,$^b$, Anurag Tripathi\,$^b$}
	\vspace{2cm} \\
	\begin{flushleft}
	\it{$^a$School of Physical Sciences, National Institute of Science Education and Research, \\ An OCC of Homi Bhaba National Institute, Jatni 752050, India \\
	$^b$Department of Physics, Indian Institute of Technology Hyderabad, Kandi, Sangareddy, Telenagana State 502284, India
	}
	\end{flushleft} 
\vspace{2cm}	
\begin{flushleft}
{\it E-mail:} \href{mailto:dragarwalphysics@gmail.com}{dragarwalphysics@gmail.com}, \href{mailto:sourav@niser.ac.in}{sourav@niser.ac.in}, \href{mailto:shrivastavadi333@gmail.com}{shrivastavadi333@gmail.com}, \href{mailto:tripathi@phy.iith.ac.in}{tripathi@phy.iith.ac.in}
\end{flushleft}
\end{center}

\begin{abstract}
\, \noindent Abstract: Infrared singularities in perturbative Quantum Chromodynamics (QCD)  are captured by the Soft function, which can be calculated efficiently in terms of multiparton webs. 
Web is a closed set of diagrams whose colour and kinematics mix through a web mixing matrix.
The web mixing matrices are computed using a well known replica trick algorithm. We present a package implemented in Mathematica to calculate these mixing matrices. Along with the package, we provide  benchmark points for several state-of-the art computations.  
\end{abstract}

\end{titlepage}
\newpage
\pagenumbering{arabic}

\noindent\textbf{Program Summary}
\vspace{1cm}

\noindent
{\em Manuscript Title:} \Cwebgen -- A tool to study colour structure of scattering amplitudes in IR limit\\
{\em Authors:} Neelima Agarwal, Sourav Pal, Aditya Srivastav, Anurag Tripathi\\
{\em Program title:} \Cwebgen\\
{\em Licensing provisions: GNU General Public License 3 (GPL)} \\
{\em Programming language:} \textsc{Wolfram Mathematica} version 11.3 or higher \\
{\em Computer(s) for which the program has been designed:} desktop PC, compute nodes/workstations\\
{\em Operating system(s) for which the program has been designed:} Linux 64bit, Mac Os, Windows\\
{\em RAM required to execute with typical data:} 8 GB and above depending on the complexity of the problem.\\
{\em Has the code been vectorized or parallelized?:} no\\
{\em Number of processors used:} any number of processors\\
{\em Supplementary material:} this article, examples
{\em Nature of problem:} Calculation of Web mixing matrices that appear in the Soft function in the study of scattering amplitudes. \\
{\em Solution method:} Starting with the colour of one diagram the code generates all the diagrams of a chosen Cweb. It then proceeds through the implementation of the replica trick to obtain Web mixing matrices. These matrices are then diagonalized and exponentiated colour factors are obtained which can be further simplified by the user separately.

\newpage
\tableofcontents

%

\section{Introduction}
Perturbative corrections of scattering amplitudes with massless gauge bosons exhibit infrared (IR) and mass (collinear) divergences.
 Over nearly a century of studies~\cite{Bloch:1937pw,Sudakov:1954sw,Yennie:1961ad,Kinoshita:1962ur,Lee:1964is,Grammer:1973db,Mueller:1979ih,Collins:1980ih,Sen:1981sd, Sen:1982bt,Korchemsky:1987wg,Korchemsky:1988hd,Magnea:1990zb,Dixon:2008gr,Gardi:2009qi,Becher:2009qa,Feige:2014wja}, several important features of these singularities have been uncovered (see~\cite{Agarwal:2021ais} for a recent review). These singularities are universal, {\it i.e.}, they do not depend on the structure of hard scattering processes. The IR divergences in a multiparton amplitude are captured by the Soft function which is given by the correlators of Wilson lines,  
\begin{align}
	{\cal S}_n \left( \zeta_i \right) \, \equiv \, \bra{0} \prod_{k = 1}^n
	\Phi \left(  \zeta_k \right) \ket{0} \, ,
	\label{genWLC}
\end{align} 
where $ \Phi \left(  \zeta \right) $ is given by,
\begin{align}
	\Phi \left(  \zeta \right) \, \equiv \, \mathcal{P} \exp \left[ {\rm i} g \!
	\int_\zeta d x \cdot {\bf A} (x) \right] \,. 
	\label{genWL}
\end{align}
 Here ${\bf A}^\mu (x) = A^\mu_a (x) \, {\bf T}^a$ is a non-abelian gauge field, 
and ${\bf T}^a$ are the generators of the gauge algebra, which can be taken to belong
to any desired representation, and $ \mathcal{P} $ denotes path ordering of the gauge fields.

\begin{figure}
	\centering
	\includegraphics[scale=0.06]{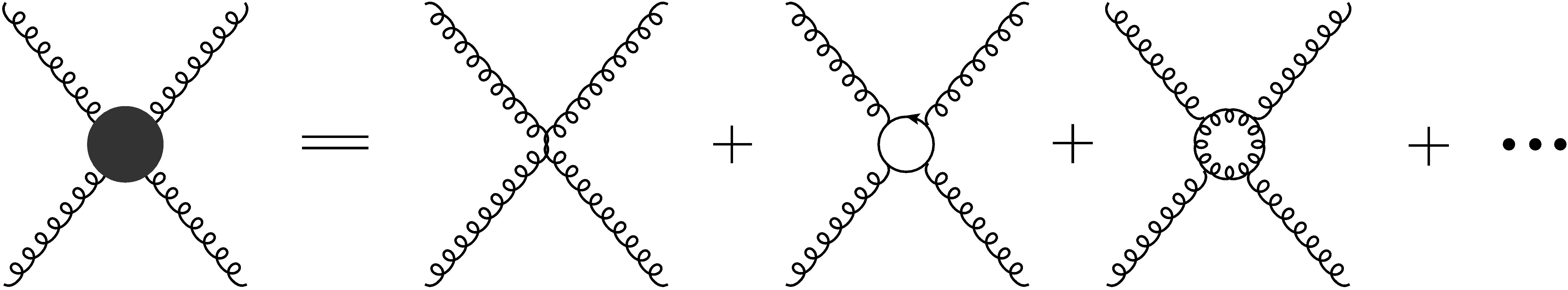}
	\caption{Perturbative expansion of a four-point gluon correlator}
	\label{fig:fourpointcorrelator}
\end{figure}

The soft function can be written as an exponential of sum of sets of diagrams: 
\begin{align}\label{eq:sumofwebs}
	\mathcal{S}_{n} =\exp\Big[ \sum W_n\Big] \,,
\end{align}
here ${W}_n $ is known as a \textit{web}.  
A web\footnote{Webs were first defined as two-line irreducible diagrams for scattering involving two Wilson lines~\cite{Sterman-1981,Gatheral,Frenkel-1984}.} in multiparton scattering amplitudes in non abelian gauge theories~\cite{Mitov:2010rp,Gardi:2010rn} is defined as the set of diagrams that is closed under the permutations of gluon attachments on each Wilson line.
%
%
\begin{report}
	Cwebs are  
related objects which are constructed out of gluon correlators  instead of gluon propagators and vertices (see four-point gluon correlator in figure~\eqref{fig:fourpointcorrelator}), these were first introduced in~\cite{Agarwal:2020nyc}. 
A Cweb is defined as a set of diagrams closed under shuffles of gluon attachments on each Wilson line. As gluon correlator has its own perturbative expansion, Cwebs are not fixed order quantities like webs.
\end{report}
%
Cwebs are especially useful for enumeration and organisation of webs at higher orders in the perturbation theory, as already shown in~\cite{Agarwal:2020nyc,Agarwal:2021him,Agarwal:2022xec,Agarwal:2023yos} for four loops and in~\cite{Mishra:2023acr} for five loops.   
 The Soft function is then given by,  
\begin{align}
	{\cal S}_n \, = \, \exp \left[ \sum_{W_n}
	\sum_{d,d' \in  W_n} \mathcal{K} (d) \, R (d, d') \, C (d')
	\right] \, .
	\label{Snwebs}
\end{align}
Here $d$ denotes a diagram in a  Cweb $W_n$, $ \mathcal{K}(d) $ and $ C(d) $ denote their kinematic and colour factors, and $R$ is called the web mixing matrix. 
\begin{report}
As Cwebs and webs are combinatorially same objects, therefore the definition of mixing matrix  for both of them remains the same and naturally they have the same properties. 
\end{report}
The mixing matrix acts on the colour of the diagrams to provide the exponentiated colour factor $\widetilde{C}(d)$ as,  
\begin{align}
	\widetilde{C} (d)  \, = \, \sum_{d'\in W_n} R(d, d') \, C(d') \, .
	\label{eq:ecf}
\end{align}
The starting point of calculating Soft function using diagrammatic exponentiation is to compute the web mixing matrices.
The web mixing matrices have three general properties: 
\begin{itemize}
	\item [(i)] Web mixing matrices are idempotent~\cite{Gardi:2010rn}, that is, $ R^2=R $. 
	\item [(ii)] The entries of $ R $ obey the zero row sum rule~\cite{Gardi:2010rn}, $ \sum_{d'} R(d,d')\,=\,0 $. 
	\item[(iii)] The mixing matrices also obey a conjectured column-sum rule~\cite{Gardi:2011yz} 
	of the form \\ 
	$\sum_{d}s(d)R(d,d')\,=\,0. $ 
\end{itemize}
For a detailed discussion on these properties and their physical implications, see refs~\cite{Agarwal:2020nyc,Agarwal:2022wyk,Agarwal:2022xec,Gardi:2010rn,Gardi:2011wa,Mishra:2023acr,White:2015wha}.

Web mixing matrices were first introduced in~\cite{Gardi:2010rn} and a replica trick algorithm was developed to calculate them. Combinatorially, application of replica trick algorithm is cumbersome and thus several attempts have been made towards calculating the mixing matrices using other methods. Use of  Posets~\cite{Dukes:2013wa,Dukes:2013gea,Dukes:2016ger}, and diagrammatic idea of Fused Webs~\cite{Agarwal:2022wyk,Agarwal:2023yos} provide two alternate methods to calculate the web mixing matrices bypassing the replica trick algorithm. Both these methods can calculate the full mixing matrices for certain classes of Cwebs. However, for complete calculation of web mixing matrices at a given loop order, employing replica trick algorithm is indispensable. 
In this paper, we address the problem of computing web mixing matrices using the well-known replica trick algorithm. 

We present a Mathematica package \texttt{\texttt{CWebGen}} that calculates the web mixing matrices by taking the expression of colour of a  single diagram of a Cweb as an  input. Our package operates independent of any external dependencies and is capable of calculating web mixing matrices at significantly higher perturbative orders.

The rest of the paper is structured as follows. In section~\ref{sec:prelim}, we explain replica trick algorithm and explain its application with an example of a two-loop Cweb. Next, in section~\ref{sec:replica}, we provide the Mathematica package \texttt{\texttt{CWebGen}} which implements the replica trick algorithm to obtain web mixing matrices discussed above. In section~\ref{installation} we provide the instructions for installation of the Mathematica package
and explain the usage of the package in \textit{integrated} and in \textit{standalone} modes. 
\begin{report}
Finally, in section~\ref{sec:sum}, we conclude the article and discuss potential future improvements for the package.	
\end{report}

\section{Preliminaries }
\label{sec:prelim}
In this section we briefly discuss the replica trick algorithm which was first developed in~\cite{Gardi:2010rn,Laenen:2008gt} and used in~\cite{Gardi:2010rn,Gardi:2013ita} to calculate three-loop web mixing matrices. Recently, we calculated four- and five-loop web mixing matrices using the same algorithm in~\cite{Agarwal:2020nyc,Agarwal:2021him,Agarwal:2022xec} and~\cite{Mishra:2023acr} respectively.
We outline the steps of the replica trick algorithm below and  provide an example on how to calculate it for a two-loop Cweb. 

\subsection{Replica Trick algorithm}
Replica trick is often used in statistical physics to calculate partition functions with very large number of particles~\cite{MezaPariVira}. It has been used for the soft gluon exponentiation~\cite{Gardi:2010rn} and recently explored for the study of soft quark exponentiation~\cite{vanBeekveld:2023liw} as well. In this section, we explain the use of replica trick in the context of soft gluon exponentiation. To start with, consider the following path integral, 
\begin{align}
	\mathcal{S}_n(\gamma_i)=\,\int \mathcal{D}A_\mu^a\,\exp(iS(A_\mu ^a)) \prod _{k=1}^n\phi_k(\gamma_k)=\exp[\mathcal{W}_n(\gamma_i)]\,,
\end{align}
where $ S(A_\mu ^a) $ is the classical action of the gauge fields and $ \phi_k $ are the Wilson lines. Note that, here $ \mathcal{W}_n $ denotes the sum of webs as described in eq.~\eqref{eq:sumofwebs}. 
In order to implement the replica trick algorithm, \( N_r \) non-interacting, identical copies of each gluon field \( A_\mu \) are introduced, meaning that each \( A_\mu \) is replaced by \( A_\mu^i \) for \( i = 1, \ldots, N_r \). For each replica, a corresponding copy of each Wilson line is also associated, effectively replacing each Wilson line with a product of \( N_r \) Wilson lines. Consequently, the path integral in the replicated theory is expressed as follows:
\begin{align}
	{\cal S}_n^{\, {\rm repl.}} \left( \gamma_i \right) \, = \,   \Big[ 
	{\cal S}_n \left( \gamma_i \right) \Big]^{N_r} \, = \, \exp \Big[ N_r \,
	{\cal W}_n (\gamma_i) \Big] \, =  \, {\bf 1} + N_r \, {\cal W}_n (\gamma_i) 
	+ {\cal O} (N_r^2) \, .
	\label{exprepl}
\end{align}
Now, using this equation, one can determine $ \mathcal{W}_n $ by calculating $ \mathcal{O}(N_r) $ terms of the Wilson line correlator in the replicated theory. The method of replicas involves five steps, which are summarized below.  
\begin{enumerate}
	\item [-] Associate a replica number to each gluon correlator in a Cweb. This number is associated to each of the colour generators of that correlator; for example a gluon attachment on the Wilson line $ k $, is represented by the colour generator $ \textbf{T}^{(i)}_k $ which belongs to a correlator with replica number $ i $.
	\item [-] In the next step one needs to find the hierarchies between the replica numbers present in a Cweb. If a Cweb has \( m \) connected pieces, the number of possible hierarchies, denoted by \( h(m) \), corresponds to the Bell number or Fubini number~\cite{IntSeq}. The first few Fubini numbers are given by $ h(m)=\{1,1,3,13,75,541\} $ for $ m= 0,1,2,3,4,5$. 
	
	\item[-] Define a replica ordering operator $ \textbf{R} $ which acts on a product of colour matrices, say $ \textbf{T}^{(i)}_k \textbf{T}^{(j)}_k $ along a given line $ k $. The action of $ \textbf{R} $ for two color matrices acting on line $ k $ is defined as
	\begin{equation}
		{\cal R}\left[{\bf T}_k^{(i)}\,{\bf T}_k^{(j)}\right]=\left\{\begin{array}{c}
			{\bf T}_k^{(i)}\,{\bf T}_k^{(j)}\quad i\leq j\\
			{\bf T}_k^{(j)}\,{\bf T}_k^{(i)}\quad i>j\end{array}\right..
		\label{repdef}
	\end{equation}
    This ordering on each line gives the replica ordered colour factor for a given diagram.
	
	\item [-]  Now one calculates the multiplicity $ M_{N_r}(h) $, which counts the number of appearances of a given hierarchy $ h $ in the presence of $ N_r $ replicas, as 
	\begin{report}
	\begin{align}
		M_{N_r}(h) \,=\, ^{N_r}{\rm C}_{n_r(h)}\,=\,\dfrac{N_r!}{n_r(h)!\,(N_r-n_r(h))!}
	\end{align}
	\end{report}
where $ n_r(h) $ is  the number of distinct replica number.
	\item [-] For a diagram $ D $, colour factor in the replicated theory is then given by, 
	\begin{align}
		C_{N_r}^{\, {\rm repl.}}  (D) \, = \, \sum_h M_{N_r} (h) \, \textbf{R} \big[ C(D) \big| h 
		\big]  \, ,
		\label{eq:expocolf}
	\end{align}
	where $ \textbf{R} \big[ C(D) \big| h \big]$ is the replica ordered colour factor of diagram $ D $, for hierarchy $ h $. Finally, the exponentiated colour factor (ECF) for diagram $ D $ is computed by extracting the coefficient of $ \mathcal{O}(N_r) $ terms of the above equation.  	
\end{enumerate} 
\subsection{A two-loop example}
In this section, we illustrate the calculation of the mixing matrix for a two-loop web. Later, in section~\ref{sec:usage} we will show how to obtain the same mixing matrix using our package. Throughout this paper, we will follow the notation of Cwebs given refs.~\cite{Agarwal:2020nyc,Agarwal:2021him,Agarwal:2022wyk,Agarwal:2022xec,Mishra:2023acr}.

\begin{figure}
	\captionsetup[subfloat]{labelformat=empty}
	\centering 
	\subfloat[][$ C_1 $]{\includegraphics[scale=0.09]{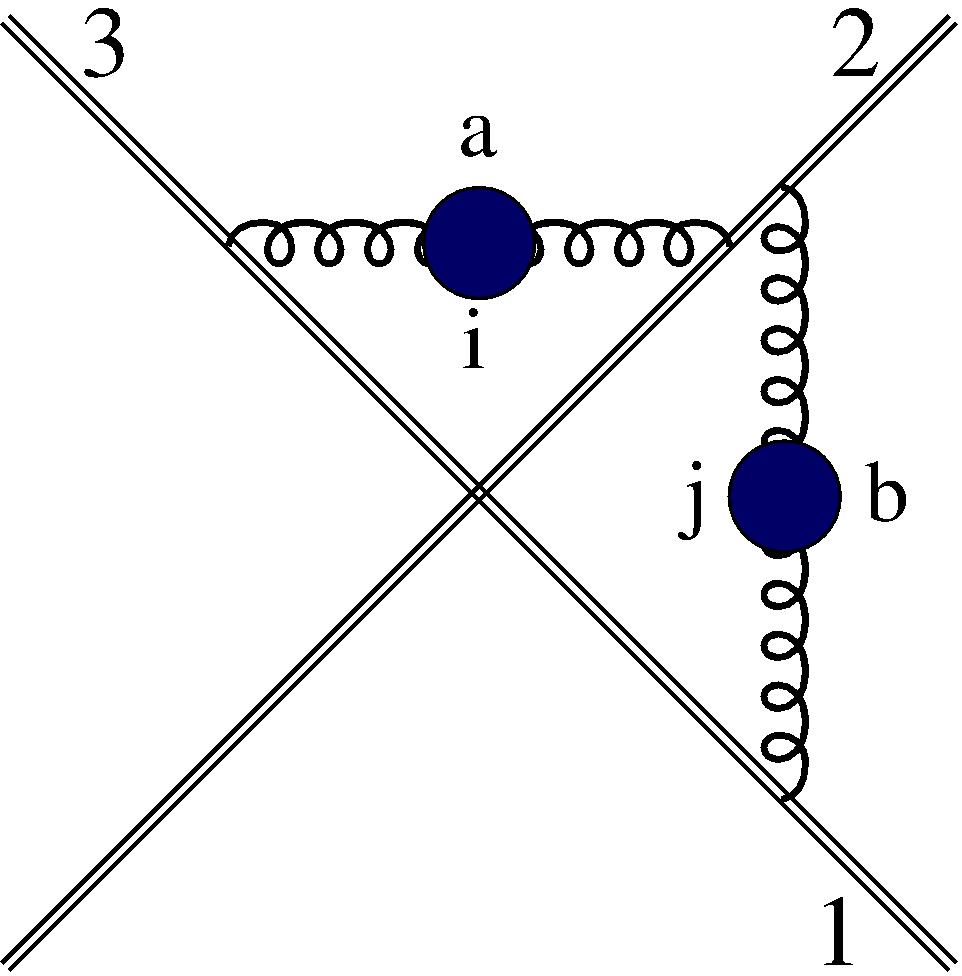} }
	\qquad \hspace{1cm}
	\subfloat[][$ C_2 $]{\includegraphics[scale=0.09]{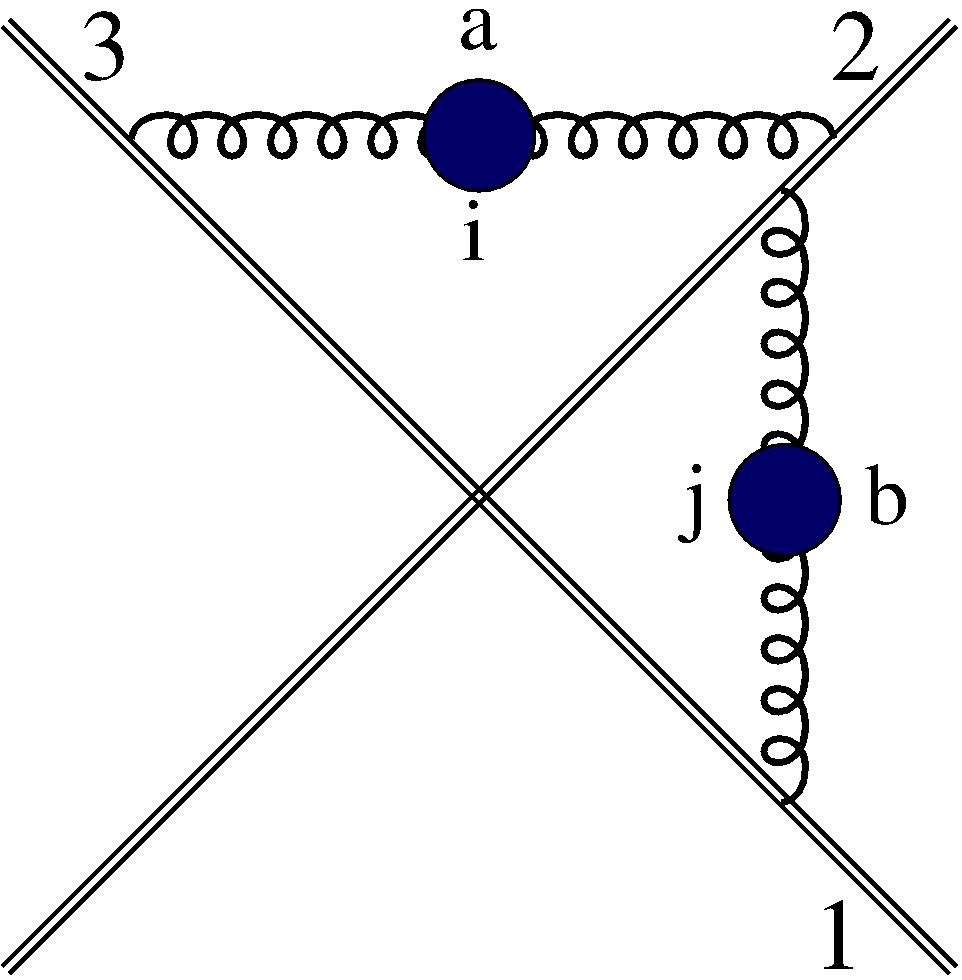} }
	\caption{Diagrams of Cweb $\text{W}\,_{3}^{(2)}\,(1,2,1)$ }
	\label{fig:Exampletwoloop-Web}
\end{figure}

We choose the Cweb \( \mathrm{W}^{(2)}_3 (1,2,1) \) to explain the application of the replica trick in calculating the mixing matrices. The first step in the calculation is to generate the set of diagrams that remain closed under shuffle of the attachments on the Wilson lines. For this specific Cweb, the closed set contains two diagrams, as shown in Fig.~\eqref{fig:Exampletwoloop-Web}. In the original (unreplicated) theory, the amplitudes of these two diagrams are given by
\begin{align}
A_1 = \mathcal{K}_1 C_1, \qquad A_2 = \mathcal{K}_2 C_2,
\end{align}
where \( \mathcal{K}_i \) and \( C_i \) represent the kinematic and color factors of diagram \( i \), respectively. In the replicated theory, the color factors of these diagrams are modified to \( \widetilde{C}_i \). Our goal is to express the color factors in the replicated theory in terms of the color factors of the diagrams in the original (unreplicated) theory.

As mentioned in the previous section, we assign replica numbers to each of the two correlators. For the case in hand, we assign replica $i$ and $j$ to the two two-point gluon correlators. The action of replica ordering operator $\bf R $ only depends on whether the replica numbers are greater, equal or smaller than each other.

The orderings of the replica numbers are hierarchies, which in this case are given by: $i=j, i>j, i<j$. For diagram $C_1$, consider the case where $i=j$, the action of $\bf R $ does not change the colour of the diagram and the colour in the replicated theory remains $C_1$. For $ i>j $, the gluons are ordered according to their replica numbers, thus $ {\bf R}(C_1|i>j)=C_2 $, similarly for $ i<j $, $ {\bf R}(C_1|i<j)=C_1 $. Similar analysis can be performed for $ C_2 $ as shown in table~\eqref{tab:rtex}. 

The next object  $M_{N_r} (h)$ can be obtained simply by summing over number of appearances of a particular hierarchy in the presence of $ N_r $ replicas. For $ i=j $, this is simply given by $ N_r $, and for $ i\neq j $, this number is given by $^{N_{r}}\rm C_2$. Then following eq.~\eqref{eq:expocolf}, the exponentiated colour factors are obtained by multiplying the $ \mathcal{O} (N_r) $ coefficient with the replica ordered colour factor. These steps are summarised in table~\eqref{tab:rtex}.    
The ECFs for this Cweb are then given by, 
\begin{align}
	\widetilde{C}_1\,&=\, \frac{1}{2} \left(C_1-C_2\right) \nonumber \\
	\widetilde{C}_2\,&=\, \frac{1}{2} \left(C_2-C_1\right)\,.
\end{align}
The corresponding mixing matrix is then given by, 
\begin{align}
	\begin{split}
		R=\,\frac{1}{2} \left(
		\begin{array}{cc}
			1 & -1 \\
			-1 & 1	
		\end{array}
		\right)\,. 
	\end{split}
\end{align} 
\begin{table}[t]
	\begin{center}
		\begin{tabular}{|c|c|c|c|c|}
			\hline
			Diagrams & hierarchy & $\mathbf{R} (C_i|h)$ & $M_{N_r}$ & $\mathcal{O} (N_r)$ \\ \hline 
			$\bf C_1$ & $i=j$ & $\bf C_1$ & $^{N_{r}}\rm C_1$ & 1\\ 	\cline{2-5}
			& $i>j$ & $\bf C_2$ &  $^{N_{r}}\rm C_2$ & $-\frac{1}{2}$\\ \cline{2-5} 
			& $i<j$ & $\bf C_1$ &  $^{N_{r}}\rm C_2$ & $-\frac{1}{2}$\\ \hline  
			$\bf C_2$ & $i=j$ & $\bf C_2$ & $^{N_{r}}\rm C_1$ & 1\\ 	\cline{2-5}
			& $i>j$ & $\bf C_2$ &  $^{N_{r}}\rm C_2$ & $-\frac{1}{2}$\\ \cline{2-5} 
			& $i<j$ & $\bf C_1$ &  $^{N_{r}}\rm C_2$ & $-\frac{1}{2}$\\ \hline  
			
		\end{tabular}
	\end{center}
	\caption{Replica trick analysis of $\rm W_3^{(2)} (1,2,1) $.\label{tab:rtex}}
\end{table}

\noindent In our package \texttt{\texttt{CWebGen}}, we automate the process described above and generate table~\eqref{tab:rtex} for each web. The next section provides a detailed explanation of the implementation of these steps in various subroutines of \texttt{\texttt{CWebGen}}.

\section{Replica Trick -- Implementation}
\label{sec:replica}

\begin{report}
The web mixing matrices are combinatorial objects and  obtained using replica trick, and the steps to obtain them are given in the previous section. 
In this section we describe the subroutines of \texttt{CWebGen} that are used to calculate these matrices.
%
%
\end{report}
\begin{enumerate}
	\item To initiate the package, one needs to provide the colour factor of one of the diagrams of a web tagged with replica indices. An example of this input colour factor is presented in the next section, and in the example files. 
	\item The next step is to generate the colour factors for all the diagrams present in a Cweb. We perform this step in a subroutine \texttt{diagramsOfWeb}. This subroutine generates shuffles among the gluon attachments on each Wilson lines and provides the list of colour of diagrams of a web. 
	\item In the next step we generate hierarchies among the replica numbers. Note that hierarchies are the ordering among replica numbers associated with gluon correlators. This step is performed in a subroutine named \texttt{hierarchies}. 	To perform this step we associate positive integers to replica numbers then use tuples to generate orderings among these integers.

	\item In the next step we construct the replica ordering operator $\bf R$ and apply it on each diagram of a web to calculate the replica-ordered colour factor. We perform this step via subroutine named \texttt{repOrdColour}. Once the replica numbers are associated with the positive integers, for a given hierarchy it is then straightforward to arrange the colour on each leg of a diagram in ascending order of replica numbers. This effectively performs the action of operation of $ \bf R $.
	
	\item The next step is to calculate  $M_{N_r} (h)$ which can be obtained simply by summing over number of appearances of a particular hierarchy in the presence of $ N_r $ replicas. We need to extract $\mathcal{O}(N_r)$ terms of this quantity as mentioned in eq.~\eqref{eq:expocolf}.

		 For example, in a hierarchy $i=j, i\geq k$, $n_r=2$. For each hierarchy, we obtain the number of distinct replica variables and obtain their $\mathcal{O}(N_r)$ terms by a subroutine named \texttt{orderNCoeff}.
	\item  The above five steps complete the five rows of table~\eqref{tab:rtex}. A final function \texttt{mixingMatrix} combines all the above subroutines to calculate the mixing matrices. Once we have the mixing matrix, we diagonalise them by obtaining diagonalising matrices constructed out of the right eigenvectors of the mixing matrices. The independent ECFs are then obtained by multiplying the diagonalising matrix to the column vector constructed out of the colours of the diagrams of a web.

\end{enumerate}

\section{\texttt{\texttt{CWebGen}} - Installation and Usage}\label{installation} \label{sec:usage}
The package can be downloaded using the link,   
\begin{center}
	\href{https://github.com/souravhep/CwebGen}{https://github.com/souravhep/CwebGen}\quad.
\end{center}
Alternatively, it can also be obtained using the command 
\begin{center}
\texttt{git clone https://github.com/souravhep/CwebGen.git}\quad .
\end{center}
The package \texttt{\texttt{CWebGen}} does not need an installation and can be directly called in a Mathematica session. 
In the following subsections, we provide the details of usage and illustrate with several examples.
\subsection{Inputs}
The package can be operated in two modes:
\texttt{Integrated} and \texttt{Standalone}. As the name suggests, the integrated mode provides the mixing matrices of webs directly, whereas the Standalone mode provides step-by-step results to the user.  In  both of these modes the input format is the same.  

To initiate the package, one needs to provide the colour factor of one of the diagrams of a web tagged with replica indices.
In Catani-Seymour colour notation~\cite{Catani:1996vz}, to write down the colour of gluon one needs two indices: (i) the leg index and (ii) the SU(3) index of the gluon (non-abelian gauge boson). 
\begin{report}
Though this notation is very useful for simplifications, in \texttt{CWebGen}, 
	 the colour 
	 	 $ 		\texttt{T}\,=\, \texttt{T}\,[\texttt{Leg number}, \texttt{Replica index}, \texttt{SU(3) index}] $
	 due to a gluon attachment 
	  has three indices:
	 \begin{itemize}
		 \item [-] Leg number: the Wilson line to which gluon is attached.
		 \item  [-]Replica index: replica index associated with the gluon correlator. 
		 \item  [-]SU(3) colour index: the SU(3) colour index of the correlator.  
	 \end{itemize}
\end{report}
 For example, in a case where a gluon is attached to Wilson line $ 3 $ with SU(3) index $ a $, the colour generator is $ 	{\mathbf T}_{3}^a $.
The input form of this colour generator is \texttt{T[3,i,a]}, where $ i $ is the 
replica index. 

For more than one attachments on a Wilson line, 
the colour factor for that line includes noncommutative product of several such generators. 
The colour for a diagram is a non-commutating product of such colour generators on each line. 
Along with these, one also needs to provide the structure constants.
Since the structure constants commute, they can be provided using \texttt{Times} function of Mathematica. For example,
\begin{align}
	f^{abc}f^{cde} \quad \longrightarrow \quad \texttt{f[a,b,c]*f[c,d,e]} \,,
\end{align}
where the LHS denotes colour factors in  SU(3) , whereas the RHS is in the format required as  an input to run the package.

The replica trick algorithm only requires the information about attachments on the Wilson lines and not what is happening away from the Wilson lines.
Therefore in our package, we consider the attachments denoted by head \texttt{T} and structure constant denoted by head \texttt{f}
as two different objects. Though the structure constants do not take part in the replica trick algorithm, providing them as an input to the package is necessary to generate the correct expressions for ECFs. 

Except the \texttt{mixingMatrix[ ]} none of the other modules of \texttt{\texttt{CWebGen}} needs the structure constants.
The colour factor that we feed in for each web has both the generators and structure constants; the colour factor equals \texttt{webcolour $ \times $ ff}, where 
\begin{align}\label{eq:webcolour-f}
	\texttt{webcolour} &=\texttt{T[1,i1,a1]**T[1,i2,a2]**\dots **T[n,in,an]} \\
	\texttt{ff}&=\texttt{f[a1,b1,c1]*f[a2,b2,c2]*...*f[an,bn,cn]}\nonumber\,.
\end{align}
For the two-loop web shown in fig.~\eqref{fig:Exampletwoloop-Web}, the colour factors of the diagrams do not have any structure constants in them, the colour factors of diagram $ C_1 $ is
\begin{align}
	C_1 = {\mathbf{T}_1^b} {\mathbf{T}_2^b} {\mathbf{T}_2^a} {\mathbf{T}_3^a}\,,
\end{align}
which translates to the input for \texttt{CWebGen} as,
\begin{align}\label{eq:2loopwebdata}
	\texttt{webcolour} &=\texttt{T[1,j,b]**T[2,j,b]**T[2,i,a]**T[3,i,a]} \\
	\texttt{ff}&=1\,. \nonumber
\end{align}
%
\subsection{Operating modes}\label{sec:operatingmodes}

\texttt{\texttt{CWebGen}} offers two operational modes: \texttt{Integrated} and \texttt{Standalone}. Both of these modes take inputs in the same format as described in the previous section. 

\subsubsection*{Integrated Mode}\label{sec:integratedmode}  

In this mode of the package user needs to access only the module 
\begin{align*}
	\texttt{mixingMatrix[webcolour,ff]}\,,
\end{align*}
and supply the two colour inputs for a web as explained in the eq.~\eqref{eq:webcolour-f}. This module creates a directory \texttt{results/} containing four files whose contents are listed below:
\begin{align*}
	\texttt{diagrams.m}& \quad \text{contains a list of diagrams (colours) of the web}\\
	\texttt{R.m}& \quad \text{contains the mixing matrix and its rank,}\\
	& \quad \text{default basis for $ {R} $ is the list of diagrams provided in \texttt{diagrams.m}}\\
	\texttt{Y.m}& \quad \text{contains a diagonalizing matrix for $ R $ obtained from its right eigenvectors}\\
	\texttt{ECFs.m}& \quad \text{contains a list of exponentiated colour factors ($ YC $) of the web}	
\end{align*}
Along with these files, it also checks whether $ R $ satisfies zero row-sum rule and is idempotent, in the current session; if these hold, following will be displayed 
\begin{align*}
	\texttt{idempotence}   & =  \texttt{True}\\
	\texttt{Row Sum} & = \{0,0,\dots,0\}
\end{align*}  

Now we will illustrate this module with the two-loop web, shown in fig.~\eqref{fig:Exampletwoloop-Web} as an example.
Calling \texttt{mixingMatrix[]}
with the colour data of diagram $ C_1 $ provided in eq.~\eqref{eq:2loopwebdata}
as
\begin{align*}
	\texttt{mixingMatrix[T[1,j,b]**T[2,j,b]**T[2,i,a]**T[3,i,a],1]}\,,
\end{align*}
 creates a \texttt{results/} directory with the following files: 
\paragraph{\texttt{diagrams.m}}
\begin{center}
	\texttt{\{\;T[1,j,b]**T[2,j,b]**T[2,i,a]**T[3,i,a],\\
		\;T[1,j,b]**T[2,i,a]**T[2,j,b]**T[3,i,a]\;\}}
\end{center}
Here we see that the code correctly generates the two diagrams of this web.
\paragraph{\texttt{R.m}}
\begin{center}
	\texttt{\{1, \{\{1/2, -1/2\}, \{-1/2, 1/2\}\}\}}
\end{center}
The first entry of this list is the rank $( r )$ of the mixing matrix, and the second is an array which is the mixing matrix $ R $.
\paragraph{\texttt{Y.m}}
\begin{center}
	\texttt{\{\{-1, 1\}, \{1, 1\}\}}
\end{center}
It is an array representing the diagonalizing matrix $ Y $.
%
\paragraph{\texttt{ECFs.m}}
\begin{center}
	\texttt{\{T[1,j,b]**T[2,i,a]**T[2,j,b]**T[3,i,a]**1  
		-T[1,j,b]**T[2,j,b]**T[2,i,a]**T[3,i,a]**1\}}
\end{center}
This file contains the ECFs which is obtained using $ Y $ ( not using $ R $, for the difference  see~\cite{Agarwal:2020nyc,Agarwal:2021him}). As expected we find only one ECF in this case which is consistent with the rank of the mixing matrix.

\begin{report}
As given in eq.~\eqref{Snwebs}, the contribution of a web in the exponent of the soft function is given by 
\begin{align}
W_n =	 \mathcal{K}^T  \, R \, C \; = \;(\mathcal{K}^T Y^{-1})\,  \,Y R Y^{-1}\, \,(YC)\;=\; 
\sum_{i=1}^{r}(\mathcal{K}^T Y^{-1})_i(YC)_i\,,
\end{align}
where vector $\mathcal{K}$ has the kinematic factors of the diagrams present in a web and $ r $ is the rank of $ R $. The package \texttt{CwebGen} provides the diagonalising matrices $Y$ with which one can construct linear combination $ (\mathcal{K}^T Y^{-1}) $
which enters the exponent in eq.~\eqref{Snwebs}; explicit form of $ \mathcal{K} $ requires computation of loop integrals.
The diagonalising matrices $Y$ are extensively used in~\cite{Gardi:2010rn,Gardi:2011yz} to show that the leading divergences of each diagram in a web cancel out. Further, these matrices for all three loop webs were used in~\cite{Gardi:2010rn,Almelid:2015jia} to calculate the soft anomalous dimension. 
\end{report}
\subsubsection*{Standalone mode}
In this mode of \texttt{CWebGen} there are modules for each subroutine and they provide the associated outputs. Here, one can obtain the intermediate results of the computation.
Below we provide the list of available functions  and their functionality. 

\begin{enumerate}
	\item 	\texttt{diagramsOfWeb[webcolour]}: This function takes as input the colour factor of one of the diagrams and shuffles the generators to generate all the diagrams of the web. The output of this function is a list of all the diagrams (colour) of the web. For example, for the two loop case mentioned in previous section, executing
	\begin{align}
		\texttt{diagramsOfWeb[T[1,j,b]**T[2,j,b]**T[2,i,a]**T[3,i,a]]}
	\end{align}
	gives the following list:
	\begin{center}
		\texttt{\{\;T[1,j,b]**T[2,j,b]**T[2,i,a]**T[3,i,a],\\
			\;T[1,j,b]**T[2,i,a]**T[2,j,b]**T[3,i,a]\;\}}
	\end{center}
	
	\item 	\texttt{hierarchies[webcolour]}: The input to this function is again the same -- colour of one of the diagrams. For a web with $ n $-correlators it provides all the possible orderings among the correlator indices (replica indices). Supplying the colour of the diagram $ C_1 $ of the above two-loop web to this function gives  
	\begin{align*}
		\texttt{hierarchies[T[1,j,b]**T[2,j,b]**T[2,i,a]**T[3,i,a]]} = \{\{1,1\},\{1,2\},\{2,1\}\}\,.
	\end{align*}
	Here $ j,i $ are the replica indices of the two correlators. Each sublist of the output defines an integer ordering which is understood as the ordering of replica indices
	\begin{align*}
		\{\{1,1\},\{1,2\},\{2,1\}\}\;=\;\{\{j=i\},\{j<i\},\{j>i\}\}\,.
	\end{align*}
	\item 	\texttt{repOrdColour[colour,hierarchy]}: This function is used to calculate the replica ordered colour factor for each diagram of the web subjected to a given hierarchy. For example, to calculate replica ordered colour factor for second diagram ($ C_2 $) with $ \{j>i\} $, executing
	\begin{center}
		\texttt{repOrdColour[T[1,j,b]**T[2,i,a]**T[2,j,b]**T[3,i,a],\{2,1\}]}\,,
	\end{center}
	 we get 
	 \begin{align*}
	 \texttt{T[1,j,b]**T[2,j,b]**T[2,i,a]**T[3,i,a]},
	 \end{align*}
 which is the colour of the first diagram ($ C_1 $).
	
	\item 	\texttt{orderNCoeff[hierarchy]}: 
	It provides $ \mathcal{O}(N_r) $ coefficient of $M_{N_r} (h)$ for a given hierarchy $ h $. For the case illustrated above, i.e. $ \{j>i\} $ we get,
	\begin{center}
		\texttt{orderNCoeff$ \left[\{2,1\}\right] \;=\; -\,\dfrac{1}{2} $}\,.
	\end{center}
\end{enumerate}

\subsection{Examples}

Along with the package we provide a directory of examples for Cwebs appearing upto six loops. The examples can be downloaded from 
\begin{center}
\href{https://github.com/souravhep/CWebGen/tree/main/examples}{https://github.com/souravhep/CWebGen/tree/main/examples}\quad .
\end{center}

\begin{figure}
	\centering
	\includegraphics[scale=0.15]{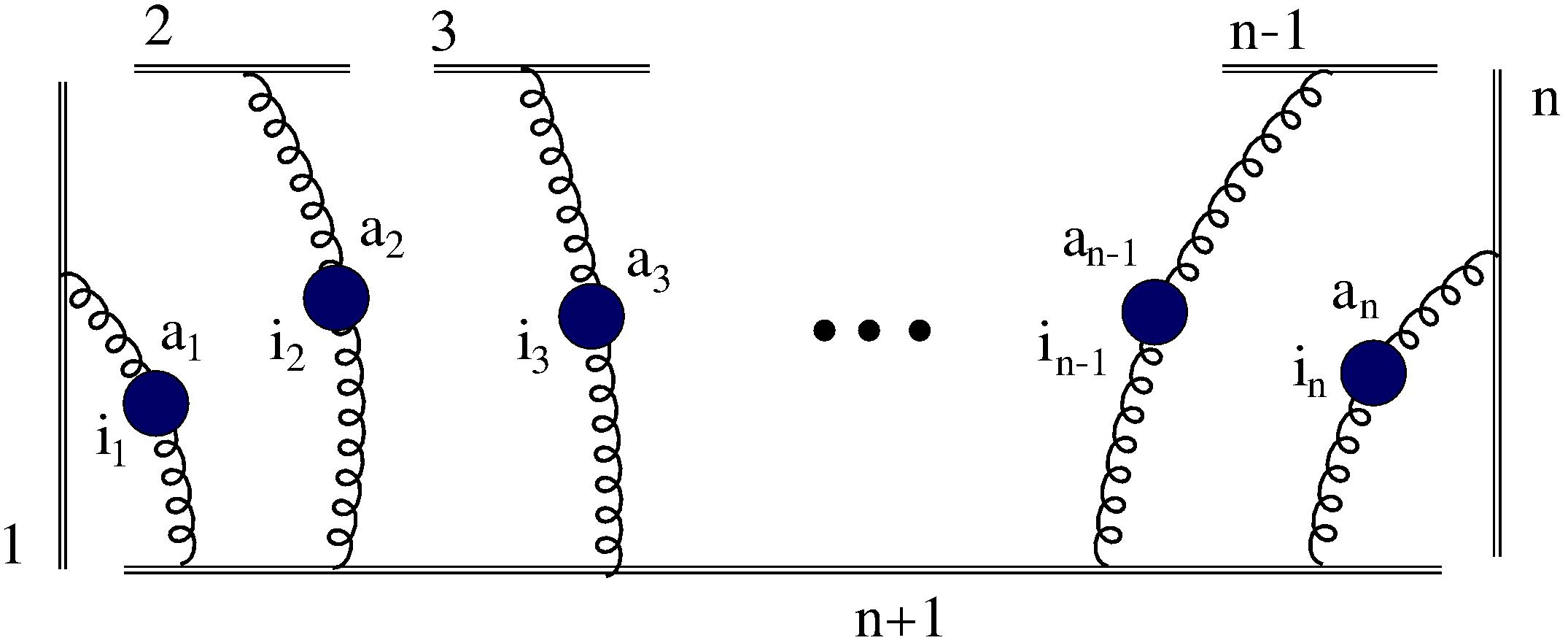}
	\caption{A diagram of a Cweb $\rm W^{(n)}_{n+1} (1,1,1,\dots,n) $}
	\label{fig:wn}
\end{figure}

\subsection{Benchmarks}
To test the efficiency of our code, we choose a class of Cwebs $ \text{W}^{(n)}_{n+1} (1,1,\ldots n) $, shown in fig.~\eqref{fig:wn} whose mixing matrices have dimension $ n! $, where $ n $ is the perturbative order. Further, the Cwebs of this class have the largest dimension of mixing matrices at a given perturbative order connecting maximum number of lines. We run a benchmark test on the computation time for the two-loop Cweb of this category. We also provide the computation times for the Cwebs of this category for three and four loops, which are already present in the literature. To test the capability of our code \texttt{\texttt{CWebGen}} in handling higher orders, we have computed mixing matrices of this category appearing at five and six loops. The computation times of all these mixing matrices are provided in table~\eqref{tab:benchmark}. 
It is evident that \texttt{\texttt{CWebGen}} is well-suited for handling higher loop computations.
\begin{table}[h]
	\begin{center}
		\begin{tabular}{|c|c|c|}
			\hline 
			loops &	Dimension of $ R $ & Computation time  \\
			\hline  
			2 & 2 & 0.021 s\\
			\hline 
			3 & 6 & 0.035 s\\
			\hline 
			4     & 24        &  0.407 s \\ 
			\hline 
			5     & 120       &   24.074 s\\
			\hline
			6     & 720       &  4.64 h \\
			\hline
		\end{tabular}
	\end{center}
	\caption{Benchmark tests of \texttt{\texttt{CWebGen}}. All these tests are performed in a computer with Intel(R) Core(TM) i7-8700 CPU @ 3.20GHz and 40GB RAM.\label{tab:benchmark}}
\end{table}

\subsection{Limitations}

While \texttt{\texttt{CWebGen}} is capable of calculating the mixing matrices for the majority of webs at a given perturbative order, it does have certain limitations.
A manual intervention is required to compute the mixing matrix of a Cweb that has identical copies of two or more correlators attached to the same set of Wilson lines in an identical fashion. 
The shuffle of these correlators leads to multiple duplicates of each diagram within a Cweb, which must be discarded before applying the replica trick algorithm. 
Since the replica trick for these Cwebs has to be applied only to a subset of the shuffles that remain after removing the duplicates our package cannot be used to generate mixing matrices for them. The complexities of applying the replica trick to this class of Cwebs are discussed in more detail in section 3 of~\cite{Agarwal:2021him}. However, one can obtain the mixing matrices for these Cwebs by manually identifying and removing the multiple duplicates of the diagrams and then applying the replica trick by using several functions of \texttt{CWebGen} in the Standalone mode. 

\begin{report}
 We emphasise that the concept of mixing matrices remains unchanged for SU($ N_C $) gauge theory and the mixing matrices for any given SU($ N_C $) theory can be obtained using this package. This is because mixing matrices are fundamentally combinatorial in nature and do not depend on the specific gauge group. 
\end{report}

\section{Summary}
\label{sec:sum}
The Soft function in perturbative QCD can be computed 
in terms of 
multiparton webs. A Web is a set of diagrams whose kinematic and colour factors mix via the web mixing matrix. A replica trick-based algorithm is crucial for calculating these mixing matrices at a given perturbative order. In this article, we introduce and describe a Mathematica package \texttt{\texttt{CWebGen}}, designed to efficiently compute web mixing matrices. Along with instructions for using the package, we provide an example to demonstrate its working. Except for webs described in the previous section, \texttt{\texttt{CWebGen}} can compute web mixing matrices at any loop order. We have tested the package for webs up to six loops and included the benchmark points. We anticipate this package will be valuable for computing web mixing matrices at higher loop orders. The future goal of the package will be to integrate it with diagram generators such as QGRAF~\cite{Nogueira:1991ex}, FeynArts~\cite{Hahn:2000kx} to generate all the Cwebs at a given perturbative order and
provide 
the colour factors of the diagrams in the required form. 
   
\section*{Acknowledgements}
NA, SP, AT would like to thank Lorenzo Magnea for collaborating on the projects that necessitated the construction of the first working version of this package. The research of SP is supported by the SERB Grant SRG/2023/000591. AS would like to thank
CSIR, Govt. of India, for an SRF fellowship (09/1001(0075)/2020-EMR-I). We would also like to thank the anonymous referees whose suggestions and comments have improved the manuscript.
\bibliographystyle{bibstyle}
\bibliography{package}

\providecommand{\href}[2]{#2}\begingroup\raggedright\begin{thebibliography}{10}

\bibitem{Bloch:1937pw}
F.~Bloch and A.~Nordsieck, \emph{{Note on the Radiation Field of the
  electron}}, \href{https://doi.org/10.1103/PhysRev.52.54}{\emph{Phys. Rev.}
  {\bfseries 52} (1937) 54}.

\bibitem{Sudakov:1954sw}
V.~V. Sudakov, \emph{{Vertex parts at very high-energies in quantum
  electrodynamics}}, {\emph{Sov. Phys. JETP} {\bfseries 3} (1956) 65}.

\bibitem{Yennie:1961ad}
D.~R. Yennie, S.~C. Frautschi and H.~Suura, \emph{{The infrared divergence
  phenomena and high-energy processes}},
  \href{https://doi.org/10.1016/0003-4916(61)90151-8}{\emph{Annals Phys.}
  {\bfseries 13} (1961) 379}.

\bibitem{Kinoshita:1962ur}
T.~Kinoshita, \emph{{Mass singularities of Feynman amplitudes}},
  \href{https://doi.org/10.1063/1.1724268}{\emph{J. Math. Phys.} {\bfseries 3}
  (1962) 650}.

\bibitem{Lee:1964is}
T.~D. Lee and M.~Nauenberg, \emph{{Degenerate Systems and Mass Singularities}},
  \href{https://doi.org/10.1103/PhysRev.133.B1549}{\emph{Phys. Rev.} {\bfseries
  133} (1964) B1549}.

\bibitem{Grammer:1973db}
G.~Grammer, Jr. and D.~R. Yennie, \emph{{Improved treatment for the infrared
  divergence problem in quantum electrodynamics}},
  \href{https://doi.org/10.1103/PhysRevD.8.4332}{\emph{Phys. Rev.} {\bfseries
  D8} (1973) 4332}.

\bibitem{Mueller:1979ih}
A.~H. Mueller, \emph{{On the Asymptotic Behavior of the Sudakov Form-factor}},
  \href{https://doi.org/10.1103/PhysRevD.20.2037}{\emph{Phys. Rev.} {\bfseries
  D20} (1979) 2037}.

\bibitem{Collins:1980ih}
J.~C. Collins, \emph{{Algorithm to Compute Corrections to the Sudakov
  Form-factor}}, \href{https://doi.org/10.1103/PhysRevD.22.1478}{\emph{Phys.
  Rev.} {\bfseries D22} (1980) 1478}.

\bibitem{Sen:1981sd}
A.~Sen, \emph{{Asymptotic Behavior of the Sudakov Form-Factor in QCD}},
  \href{https://doi.org/10.1103/PhysRevD.24.3281}{\emph{Phys. Rev.} {\bfseries
  D24} (1981) 3281}.

\bibitem{Sen:1982bt}
A.~Sen, \emph{{Asymptotic Behavior of the Wide Angle On-Shell Quark Scattering
  Amplitudes in Nonabelian Gauge Theories}},
  \href{https://doi.org/10.1103/PhysRevD.28.860}{\emph{Phys. Rev.} {\bfseries
  D28} (1983) 860}.

\bibitem{Korchemsky:1987wg}
G.~P. Korchemsky and A.~V. Radyushkin, \emph{{Renormalization of the Wilson
  Loops Beyond the Leading Order}},
  \href{https://doi.org/10.1016/0550-3213(87)90277-X}{\emph{Nucl. Phys.}
  {\bfseries B283} (1987) 342}.

\bibitem{Korchemsky:1988hd}
G.~P. Korchemsky, \emph{{Sudakov Form-factor in {QCD}}},
  \href{https://doi.org/10.1016/0370-2693(89)90799-5}{\emph{Phys. Lett.}
  {\bfseries B220} (1989) 629}.

\bibitem{Magnea:1990zb}
L.~Magnea and G.~F. Sterman, \emph{{Analytic continuation of the Sudakov
  form-factor in QCD}},
  \href{https://doi.org/10.1103/PhysRevD.42.4222}{\emph{Phys. Rev.} {\bfseries
  D42} (1990) 4222}.

\bibitem{Dixon:2008gr}
L.~J. Dixon, L.~Magnea and G.~F. Sterman, \emph{{Universal structure of
  subleading infrared poles in gauge theory amplitudes}},
  \href{https://doi.org/10.1088/1126-6708/2008/08/022}{\emph{JHEP} {\bfseries
  08} (2008) 022} [\href{https://arxiv.org/abs/0805.3515}{{\ttfamily
  0805.3515}}].

\bibitem{Gardi:2009qi}
E.~Gardi and L.~Magnea, \emph{{Factorization constraints for soft anomalous
  dimensions in QCD scattering amplitudes}},
  \href{https://doi.org/10.1088/1126-6708/2009/03/079}{\emph{JHEP} {\bfseries
  03} (2009) 079} [\href{https://arxiv.org/abs/0901.1091}{{\ttfamily
  0901.1091}}].

\bibitem{Becher:2009qa}
T.~Becher and M.~Neubert, \emph{{On the Structure of Infrared Singularities of
  Gauge-Theory Amplitudes}},
  \href{https://doi.org/10.1088/1126-6708/2009/06/081,
  10.1007/JHEP11(2013)024}{\emph{JHEP} {\bfseries 06} (2009) 081}
  [\href{https://arxiv.org/abs/0903.1126}{{\ttfamily 0903.1126}}].

\bibitem{Feige:2014wja}
I.~Feige and M.~D. Schwartz, \emph{{Hard-Soft-Collinear Factorization to All
  Orders}}, \href{https://doi.org/10.1103/PhysRevD.90.105020}{\emph{Phys. Rev.}
  {\bfseries D90} (2014) 105020}
  [\href{https://arxiv.org/abs/1403.6472}{{\ttfamily 1403.6472}}].

\bibitem{Agarwal:2021ais}
N.~Agarwal, L.~Magnea, C.~Signorile-Signorile and A.~Tripathi, \emph{{The
  infrared structure of perturbative gauge theories}},
  \href{https://doi.org/10.1016/j.physrep.2022.10.001}{\emph{Phys. Rept.}
  {\bfseries 994} (2023) 1} [\href{https://arxiv.org/abs/2112.07099}{{\ttfamily
  2112.07099}}].

\bibitem{Sterman-1981}
G.~F. Sterman, \emph{{Infrared divergences in perturbative QCD}},
  \href{https://doi.org/10.1063/1.33099}{\emph{AIP Conf. Proc.} {\bfseries 74}
  (1981) 22}.

\bibitem{Gatheral}
J.~G.~M. Gatheral, \emph{{Exponentiation of Eikonal Cross-sections in
  Nonabelian Gauge Theories}},
  \href{https://doi.org/10.1016/0370-2693(83)90112-0}{\emph{Phys. Lett.}
  {\bfseries 133B} (1983) 90}.

\bibitem{Frenkel-1984}
J.~Frenkel and J.~C. Taylor, \emph{{Nonabelian eikonal exponentiation}},
  \href{https://doi.org/10.1016/0550-3213(84)90294-3}{\emph{Nucl. Phys.}
  {\bfseries B246} (1984) 231}.

\bibitem{Mitov:2010rp}
A.~Mitov, G.~Sterman and I.~Sung, \emph{{Diagrammatic Exponentiation for
  Products of Wilson Lines}},
  \href{https://doi.org/10.1103/PhysRevD.82.096010}{\emph{Phys. Rev.}
  {\bfseries D82} (2010) 096010}
  [\href{https://arxiv.org/abs/1008.0099}{{\ttfamily 1008.0099}}].

\bibitem{Gardi:2010rn}
E.~Gardi, E.~Laenen, G.~Stavenga and C.~D. White, \emph{{Webs in multiparton
  scattering using the replica trick}},
  \href{https://doi.org/10.1007/JHEP11(2010)155}{\emph{JHEP} {\bfseries 11}
  (2010) 155} [\href{https://arxiv.org/abs/1008.0098}{{\ttfamily 1008.0098}}].

\bibitem{Agarwal:2020nyc}
N.~Agarwal, A.~Danish, L.~Magnea, S.~Pal and A.~Tripathi, \emph{{Multiparton
  webs beyond three loops}},
  \href{https://doi.org/10.1007/JHEP05(2020)128}{\emph{JHEP} {\bfseries 05}
  (2020) 128} [\href{https://arxiv.org/abs/2003.09714}{{\ttfamily
  2003.09714}}].

\bibitem{Agarwal:2021him}
N.~Agarwal, L.~Magnea, S.~Pal and A.~Tripathi, \emph{{Cwebs beyond three loops
  in multiparton amplitudes}},
  \href{https://doi.org/10.1007/JHEP03(2021)188}{\emph{JHEP} {\bfseries 03}
  (2021) 188} [\href{https://arxiv.org/abs/2102.03598}{{\ttfamily
  2102.03598}}].

\bibitem{Agarwal:2022xec}
N.~Agarwal, S.~Pal, A.~Srivastav and A.~Tripathi, \emph{{Deciphering colour
  building blocks of massive multiparton amplitudes at 4-loops and beyond}},
  \href{https://doi.org/10.1007/JHEP02(2023)258}{\emph{JHEP} {\bfseries 02}
  (2023) 258} [\href{https://arxiv.org/abs/2212.06610}{{\ttfamily
  2212.06610}}].

\bibitem{Agarwal:2023yos}
N.~Agarwal, S.~Pal, A.~Srivastav and A.~Tripathi, \emph{{Correlator webs of
  massive multiparton amplitudes at four loops: A study of boomerang webs}},
  \href{https://doi.org/10.1103/PhysRevD.109.094038}{\emph{Phys. Rev. D}
  {\bfseries 109} (2024) 094038}
  [\href{https://arxiv.org/abs/2307.15924}{{\ttfamily 2307.15924}}].

\bibitem{Mishra:2023acr}
S.~Mishra, S.~Pal, A.~Srivastav and A.~Tripathi, \emph{{Multiparton Cwebs at
  five loops}}, \href{https://doi.org/10.1007/JHEP07(2024)078}{\emph{JHEP}
  {\bfseries 07} (2024) 078}
  [\href{https://arxiv.org/abs/2305.17452}{{\ttfamily 2305.17452}}].

\bibitem{Gardi:2011yz}
E.~Gardi, J.~M. Smillie and C.~D. White, \emph{{On the renormalization of
  multiparton webs}},
  \href{https://doi.org/10.1007/JHEP09(2011)114}{\emph{JHEP} {\bfseries 09}
  (2011) 114} [\href{https://arxiv.org/abs/1108.1357}{{\ttfamily 1108.1357}}].

\bibitem{Agarwal:2022wyk}
N.~Agarwal, S.~Pal, A.~Srivastav and A.~Tripathi, \emph{{Building blocks of
  Cwebs in multiparton scattering amplitudes}},
  \href{https://doi.org/10.1007/JHEP06(2022)020}{\emph{JHEP} {\bfseries 06}
  (2022) 020} [\href{https://arxiv.org/abs/2204.05936}{{\ttfamily
  2204.05936}}].

\bibitem{Gardi:2011wa}
E.~Gardi and C.~D. White, \emph{{General properties of multiparton webs: Proofs
  from combinatorics}},
  \href{https://doi.org/10.1007/JHEP03(2011)079}{\emph{JHEP} {\bfseries 03}
  (2011) 079} [\href{https://arxiv.org/abs/1102.0756}{{\ttfamily 1102.0756}}].

\bibitem{White:2015wha}
C.~D. White, \emph{{An Introduction to Webs}},
  \href{https://doi.org/10.1088/0954-3899/43/3/033002}{\emph{J. Phys. G}
  {\bfseries 43} (2016) 033002}
  [\href{https://arxiv.org/abs/1507.02167}{{\ttfamily 1507.02167}}].

\bibitem{Dukes:2013wa}
M.~Dukes, E.~Gardi, E.~Steingrimsson and C.~D. White, \emph{{Web worlds,
  web-colouring matrices, and web-mixing matrices}},
  \href{https://doi.org/10.1016/j.jcta.2013.02.001}{\emph{J. Comb. Theory Ser.}
  {\bfseries A120} (2013) 1012}
  [\href{https://arxiv.org/abs/1301.6576}{{\ttfamily 1301.6576}}].

\bibitem{Dukes:2013gea}
M.~Dukes, E.~Gardi, H.~McAslan, D.~J. Scott and C.~D. White, \emph{{Webs and
  Posets}}, \href{https://doi.org/10.1007/JHEP01(2014)024}{\emph{JHEP}
  {\bfseries 01} (2014) 024} [\href{https://arxiv.org/abs/1310.3127}{{\ttfamily
  1310.3127}}].

\bibitem{Dukes:2016ger}
M.~Dukes and C.~D. White, \emph{{Web matrices: structural properties and
  generating combinatorial identities}},
  \href{https://arxiv.org/abs/1603.01589}{{\ttfamily 1603.01589}}.

\bibitem{Laenen:2008gt}
E.~Laenen, G.~Stavenga and C.~D. White, \emph{Path integral approach to eikonal
  and next-to-eikonal exponentiation},
  \href{https://doi.org/10.1088/1126-6708/2009/03/054}{\emph{JHEP} {\bfseries
  03} (2009) 054} [\href{https://arxiv.org/abs/0811.2067}{{\ttfamily
  0811.2067}}].

\bibitem{Gardi:2013ita}
E.~Gardi, J.~M. Smillie and C.~D. White, \emph{{The Non-Abelian Exponentiation
  theorem for multiple Wilson lines}},
  \href{https://doi.org/10.1007/JHEP06(2013)088}{\emph{JHEP} {\bfseries 06}
  (2013) 088} [\href{https://arxiv.org/abs/1304.7040}{{\ttfamily 1304.7040}}].

\bibitem{MezaPariVira}
M.~M\'ezard, G.~Parisi and M.~Virasoro, \emph{{Spin glass theory and beyond}},
  {\emph{World Scientific Lecture Notes in Physics} {\bfseries 9} (1987) }.

\bibitem{vanBeekveld:2023liw}
M.~van Beekveld, L.~Vernazza and C.~D. White, \emph{{Exponentiation of soft
  quark effects from the replica trick}},
  \href{https://doi.org/10.1007/JHEP07(2024)109}{\emph{JHEP} {\bfseries 07}
  (2024) 109} [\href{https://arxiv.org/abs/2312.11606}{{\ttfamily
  2312.11606}}].

\bibitem{IntSeq}
{\emph{{\tt https://oeis.org/A000670 \!\!}} }.

\bibitem{Catani:1996vz}
S.~Catani and M.~H. Seymour, \emph{{A General algorithm for calculating jet
  cross-sections in NLO QCD}},
  \href{https://doi.org/10.1016/S0550-3213(96)00589-5}{\emph{Nucl. Phys. B}
  {\bfseries 485} (1997) 291}
  [\href{https://arxiv.org/abs/hep-ph/9605323}{{\ttfamily hep-ph/9605323}}].

\bibitem{Almelid:2015jia}
O.~Almelid, C.~Duhr and E.~Gardi, \emph{{Three-loop corrections to the soft
  anomalous dimension in multileg scattering}},
  \href{https://doi.org/10.1103/PhysRevLett.117.172002}{\emph{Phys. Rev. Lett.}
  {\bfseries 117} (2016) 172002}
  [\href{https://arxiv.org/abs/1507.00047}{{\ttfamily 1507.00047}}].

\bibitem{Nogueira:1991ex}
P.~Nogueira, \emph{{Automatic Feynman graph generation}},
  \href{https://doi.org/10.1006/jcph.1993.1074}{\emph{J. Comput. Phys.}
  {\bfseries 105} (1993) 279}.

\bibitem{Hahn:2000kx}
T.~Hahn, \emph{{Generating Feynman diagrams and amplitudes with FeynArts 3}},
  \href{https://doi.org/10.1016/S0010-4655(01)00290-9}{\emph{Comput. Phys.
  Commun.} {\bfseries 140} (2001) 418}
  [\href{https://arxiv.org/abs/hep-ph/0012260}{{\ttfamily hep-ph/0012260}}].

\end{thebibliography}\endgroup
\end{document}